\documentclass[floatfix,reprint,aps,pra,superscriptaddress
]{revtex4-2}

\usepackage{amsmath}
\usepackage{amssymb}
\usepackage{amsfonts}
\usepackage{mathrsfs}
\usepackage{amsthm}
\usepackage{mathtools}
\usepackage{braket}
\usepackage{physics}
\usepackage{nicematrix}
\usepackage{bm} 

\theoremstyle{definition}

\usepackage[caption=false]{subfig}
\usepackage{graphicx}
\usepackage{dcolumn}
\usepackage{bm}
\usepackage{appendix}
\usepackage{amsfonts}
\usepackage[normalem]{ulem}
\usepackage{xcolor}
\usepackage{booktabs}
\usepackage{comment}

\begin{document}

\title{Optimal complexity of parameterized quantum circuits}

\author{Guilherme Ilário Correr}
\email{guilhermecorrer27@gmail.com}
\affiliation{
 Instituto de Física de São Carlos, Universidade de São Paulo, CP 369, 13560-970, São Carlos, SP, Brasil.}
\author{Pedro C. Azado}
\affiliation{
 Instituto de Física de São Carlos, Universidade de São Paulo, CP 369, 13560-970, São Carlos, SP, Brasil.}
\author{Diogo O. Soares-Pinto}
\affiliation{
 Instituto de Física de São Carlos, Universidade de São Paulo, CP 369, 13560-970, São Carlos, SP, Brasil.}
\author{Gabriel Carlo}
\email{carlo@tandar.cnea.gov.ar}
\affiliation{Comisión Nacional de Energía Atómica, CONICET, Departamento de Física, Av. del Libertador 8250, 1429 Buenos Aires, Argentina}

\begin{abstract}
Parameterized quantum circuits play a key role for the development of quantum variational algorithms in the realm of the NISQ era. Knowing their actual capability of performing different kinds of tasks is then of the utmost importance. By comparing them with a prototypical class of universal random circuits we have found that their approach to the asymptotic complexity defined by the Haar measure is faster, needing less gates to reach it. Topology has been revealed crucial for this. The majorization criterion has proven as a relevant complementary tool to the expressibility and the mean entanglement.   
\end{abstract}

\maketitle

\section{\label{sec:introduction}Introduction}
The classification of quantum random circuits according to their complexity has become an active 
area of research. On the one hand, random quantum circuits are important simulators of quantum dynamics, being of fundamental importance to generate approximations of Haar random unitaries \cite{randomcircuits_emerson_lloyd} and for the understanding of different kinds of many-body dynamics \cite{lloyd_entanglement_complexity_scrambling, randomcircuits_review, Ippoliti2022solvablemodelofdeep}. Often, these applications rely on highly complex quantum circuits to achieve specific tasks. On the other hand, it is possible to say that the main reason behind this complexity characterization is that with the advent of the so-called NISQ 
devices novel platforms have emerged as potential probes to test quantum advantage. Knowing which are the best architectures for implementing different quantum protocols like, for example, variational quantum algorithms, is also highly relevant \cite{peruzzo2014variational, McClean_Ucoupledcluster, HEA_superconductors_smallmolecules, practicalimplementation_HEA}. In this context, parameterized quantum circuits play a central role 
for the development of efficient quantum algorithms. However, their characteristics are not fully understood, even more when considered outside the scope of variational algorithms. In this sense, the study of their complexity growth is highly interesting to investigate the possibilities of applications these circuits based on NISQ devices can achieve. 

There are several measures of complexity for quantum circuits \cite{yao1993quantum, eisert2021entangling, bu2022statistical, haferkamp2022linear}. Some of them are based on a comparison between the uniform and invariant measure over the space of the group, the Haar measure, and the random unitaries generated by the circuit. A recently introduced measure of this kind, the majorization criterion \cite{major1}, is based on the fluctuations of Lorenz curves. These curves are defined by the cumulants of a given ordered distribution. By comparing these curves with the one obtained considering Haar sampled unitaries, it is possible to characterize the complexity of the circuit. Another measure that has attracted a lot of attention recently is the expressibility \cite{express_entang_capab}. This measure can be translated into the relative entropy comparing the distribution of fidelities of states generated by the random circuit with the distribution of the Haar random case. We have also considered the average entanglement that determines the average and standard deviation values for the entanglement of states generated by the circuits. This later quantity contains important information about the characteristics of the circuit and can be made suitable to future complexity assessments \cite{eisert2021entangling, Viola_parameters_of_pseudorandomcircuits, scott_baker_map, Scott_entanglement_measure, liu2018entanglement, iaconis2021quantum}. For example, random circuits close in behavior to the Haar measure, i.e. close to a $t$-design \cite{unitary_designs, polynomial_t-designs1, polynomial_t-designs2} or close to the generation of uniformly distributed states, generate average entanglement that approximate well the values obtained sampling with the Haar measure \cite{Viola_parameters_of_pseudorandomcircuits, scott_baker_map, Scott_entanglement_measure}. Therefore, it can be seen as a necessary condition and another tool to understand the evolution of entanglement correlations when circuit complexity increase.

In this work we compare these three measures for different configurations of parameterized quantum circuits and an universal class of random circuits generated by few gates and widely used in several implementations of NISQ benchmarking. As a result of our comparison, we could identify that the rate of convergence towards the optimal complexity characterized by Haar like fluctuations in the Lorenz curves is greater for the parameterized circuits given by the most connected topologies than the non-parametric class. This is consistent with the results obtained using the expressibility and the mean entanglement.

The paper is organized as follows: In Sec. \ref{sec:QC} we explain the construction of the different classes of circuits we analyze. In the following Sec. \ref{sec:quantifiers} we define the measures used to quantify the complexity of the previously defined circuits. Results are shown in Sec. \ref{sec:results}. Finally, we offer the concluding remarks in Sec. \ref{sec:conclusion}.

\section{\label{sec:QC}Quantum Circuits}

The main class of circuits that we are going to analyze are the so called parameterized quantum circuits (PQCs). These are fixed structures of parameterized gates of one and two qubits, concatenated many times to achieve different objectives \cite{express_entang_capab, Cerezo2021VqaReview}. In the context of Variational Quantum Algorithms (VQAs), these parameters are optimized by applying a classical optimization method together with a cost function that encodes the solution of a particular problem to be solved. Still, a different option to explore the possibilities for PQCs is to sample the parameters at random to obtain circuits generating ensembles of random unitaries or random states \cite{Kim_entanglementdiagnostics_VQA_randomcircuits, Kim_DarioRosa_chaosandcircuitparameters_randomcircuit,randomness_PQCs}, in a very close manner to pseudorandom circuits \cite{randomcircuits_review, randomcircuits_emerson_lloyd, Viola_parameters_of_pseudorandomcircuits}.

\begin{figure}[htb!]
    \centering
    \includegraphics[width=\columnwidth]{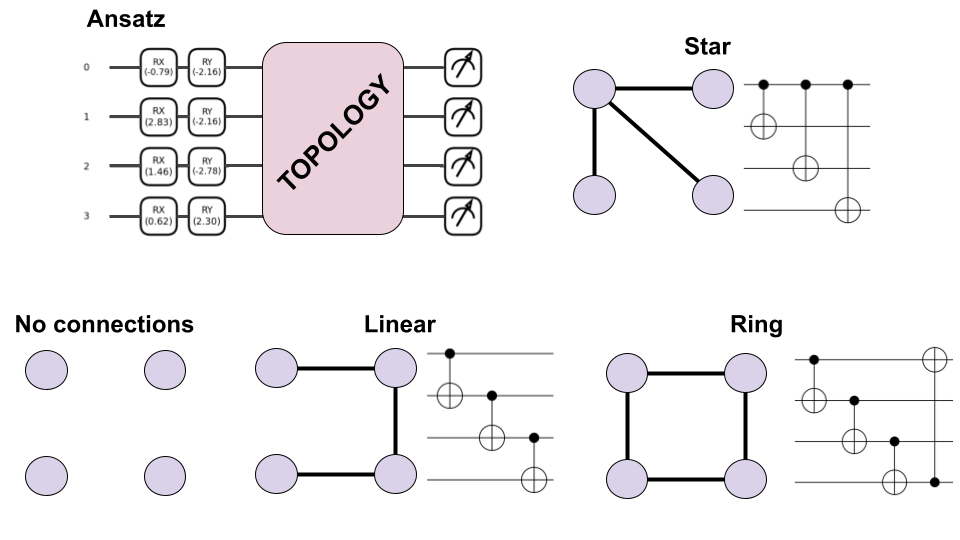}
    \caption{Circuit ansatz (fixed structure that is concatenated) and different topologies of connections considered here for $4$ qubits. The circuits are generated by changing the ``TOPOLOGY'' part to the CNOT circuit representation of the graphs. In the case of \emph{No connections}, nothing is done in the topology step.}
    \label{fig:circuits_choices}
\end{figure}

The structures of the PQCs in this work are chosen both to simplify the local parameterized operations and to match the connectivities available in the IBM quantum computers with H topology and others with similar square/rectangular topology \cite{ibm_vigo, topology_generativemodel_ibm, topology_topgen_circuitgenerator}. We considered from $4$ to $8$ qubits and from $1$ to $10$ circuit concatenations with independent parameters, called number of layers in the context of VQAs \cite{Cerezo2021VqaReview}. To compare with random circuits composed of discrete gates, the number of layers is translated to number of gates according to Table \ref{tab:numberofgates_ansatze_top}. Fig. \ref{fig:circuits_choices} presents the circuit ansätze, where RX and RY are applied to every qubit with random parameters sampled according to the uniform distribution between $0$ and $2\pi$, followed by the connections, represented as graphs and as digital circuits. CNOTs are used as the two qubits gates responsible for the connections. We present only the $4$ qubits case as an illustration. The sequence of gates RX and RY with parameters uniformly sampled is not capable of generating uniformly distributed states of one qubit when considering the $\ket{0}$ input state \cite{express_entang_capab}. Still, this choice is capable of obtaining states distributed around the Bloch sphere, and can lead to random distributed states close to the uniform distribution \cite{express_entang_capab, correr2024characterizing}.

\begin{table*}[htb]
	\caption{Total number of gates comparing topologies for the circuit structure considered in this work as a function of the number of qubits $n$ and number of layers $l$.}\label{tab:numberofgates_ansatze_top}%
		\centering{%
		\begin{tabular}{ccc}
			\toprule
			Topology & Number of CNOTs & Total number of gates \\
			\midrule \midrule
			No connections & $0$ & $(2n)l$\\
			\midrule 
			Linear & $(n-1)l$ & $(3n-1)l$ \\
			\midrule 
			Ring & $nl$ & $(3n)l$ \\
			\midrule 
			Star & $(n-1)l$ & $(3n-1)l$ \\
            \bottomrule
			\end{tabular}%
	}
\end{table*}

The circuits are executed many times with different parameters to generate the states used to calculate the quantifiers. The parameters in the parameter vector are sampled considering independent and identically distributed random variables.

For comparison purposes we are also going to study the behavior of a standard class of universal quantum circuits that is constructed by means of a few generators and which is a standard model for universal quantum computation. This is given by ${\it G3}=\{CNOT, H, T\}$, where $H$ stands for Hadamard and $T$ are $\pi/8$ phase gates.
The set {\em G3} has been proven to be universal, approximating the unitary group $U(N)$ to desired precision ~\cite{major1}. We take equal probability for each gate at a given time, and also equal probability for the qubits or pairs of qubits to apply them.

\section{\label{sec:quantifiers}Complexity Quantifiers}

\subsection{\label{subsec:expressibility}Expressibility}

The expressibility is a figure of merit proposed in the context of parameterized quantum circuits to analyse how uniformly distributed are the pure states generated by the circuit in the state space \cite{express_entang_capab}. To do so, the circuit averaged state over randomly distributed parameters is compared to the averaged state considering the uniformly distributed Haar measure \cite{express_entang_capab}
\begin{equation}
    A^{(t)}=\int_{Haar} (\ketbra{\psi}{\psi})^{\otimes t} d\psi - \int_{\Theta} (\ketbra{\phi(\boldsymbol{\theta})}{\phi(\boldsymbol{\theta})})^{\otimes t} d\boldsymbol{\theta},
\end{equation}
being $\Theta$ the space of parameter vectors given as input for the circuit, considering a particular distribution for the sampling, and $d\psi$ the uniformly distributed Haar-induced measure over pure states space \cite{how_to_generate_rmatrices_mezzadri, Measuresspaceofstates_Zyczkowski_2001}. The Hilbert-Schmidt norm is calculated for this quantity and the closer it is to $0$, the closer the circuit is to generating uniformly distributed states. This more rigorous definition compares the circuits with $t$-designs, which is a good measure to quantify how close a circuit is to a particular design order (i.e., how close the moments of the circuit are to the Haar ones up to the $t$-th moment \cite{unitary_designs, polynomial_t-designs1, chaosandcomplexity_beniyoshida}). However, a more broad and operationally meaningful quantifier based in the same notion is defined by the relative entropy computed considering the distribution of fidelities comparing two states generated by the circuit and the same distribution for Haar random states. The relative entropy or Kullback-Leibler divergence \cite{kullback1951information, wilde_shannontheory} is defined as
\begin{equation}
    D_{KL}(P||Q) = \sum_x P(x) \log\left({\frac{P(x)}{Q(x)}}\right).
\end{equation}
The distributions $P$ and $Q$ are the fidelities distributions. First, the circuit distribution is computed sampling states with different parameters $\ket{\psi(\boldsymbol{\theta})}$, $\ket{\psi(\boldsymbol{\varphi})}$ and calculating the fidelity $F(\boldsymbol{\theta}, \boldsymbol{\varphi})=\left|\braket{\psi(\boldsymbol{\theta})}{\psi(\boldsymbol{\varphi})}\right|^2$. From sampling many different states, a histogram can be built, $P_{PQC}(F)$. The sample size was $10^4$ parameter vectors and, therefore, output states, generating $5\cdot10^3$ fidelities. This histogram is then compared with the one obtained with the probability density function of fidelities for Haar random states, $PDF_{\text{Haar}}(F)=(d-1)(1-F)^{d-2}$, being $d$ the dimension of the system \cite{zyczkowski2005average}. We call this histogram $P_{\text{Haar}}(F)$. This way, to estimate how uniformly distributed are the states generated by the circuit, we compute
\begin{equation}
    \text{Expr}:=D_{KL}\left(P_{PQC}(F)||P_{\text{Haar}}(F)\right).
\end{equation}
The closer $\text{Expr}$ is to zero, the more uniformly distributed are the states generated by the circuit in the state space. Usually, it is then said that the circuit induced measure is more \textit{expressible}. To avoid misunderstandings, we are going to refer to $\text{Expr}$ as $D_{KL}$ or relative entropy, and the closer this quantity is to zero, the higher the expressibility.

\subsection{\label{lorenzcurves}Majorization criterion}

In trying to grasp the complexity of quantum circuits (and devices) a measure inspired in the majorization principle has 
been recently proposed ~\cite{major1}. By majorization we refer to a way of ordering vectors according to the distribution of their components. We can take any two vectors $ \textbf{p}, \textbf{q} \in \mathbb{R}^N $, for example. If  
\begin{align}
    \sum_{i=1}^{k} p_{i}^{\downarrow} & \leq \sum_{i=1}^{k} q_{i}^{\downarrow}, \quad 1\leq k < N, \label{eq:major_ineq}\\
    \sum_{i=1}^{N} p_{i} & = \sum_{i=1}^{N} q_{i}\label{eq:normalization},
\end{align}
where $^{\downarrow}$ stands for sorting the components in non-increasing order, then \textbf{p} is majorized by \textbf{q}. 
This is usually written as $\textbf{p} \prec \textbf{q}$, and it indicates that the components of \textbf{p} are more 
uniformly distributed than the components of \textbf{q} ~\cite{major2}. In our case the components are the probabilities associated 
to the output state vectors of a given quantum circuit (normalized). The $k$-th partial sum in Eq.~(\ref{eq:major_ineq}) is called 
the $k$-th cumulant of either $\textbf{p}$ or $\textbf{q}$ ($\mathcal{F}_p(k)$ and $\mathcal{F}_q(k)$, respectively). It is then clear that 
if $\textbf{p} \prec \textbf{q}$, then $\mathcal{F}_p(k) \leq \mathcal{F}_q(k)$ for $1\leq k < N$. The plots of $\mathcal{F}_p(k)$ and $\mathcal{F}_q(k)$ vs $k/N$ are 
the \emph{Lorenz curves} and saying that \textbf{q} majorizes \textbf{p} is equivalent to the Lorenz curve for \textbf{q} being 
always above the curve for \textbf{p}.

If we consider an ensemble of $n$-qubit random quantum circuits $\{U\}$ of a given class we can measure this class complexity by 
studying the fluctuations of the Lorenz curves. This is accomplished by uniformly sampling the corresponding circuits in order to 
make them act on an initial state given by $|0\ldots 0\rangle= |0\rangle^{\otimes n}$ and finally measuring in the computational basis. 
This gives the output distributions, $p_U (i) = \left|\bra{0\ldots 0} U \ket{i}\right|^2$, whose cumulants 
$\mathcal{F}_{p_U}(k)$ -- with $k\in\{1,\ldots,2^n\}$ are used to evaluate the fluctuations 
\begin{align}\label{eq:stdF}
    {\rm std}\,[\mathcal{F}_{p_U} (k)] = \sqrt{\langle \mathcal{F}_{p_U}^2 (k) \rangle - \langle \mathcal{F}_{p_U}(k) \rangle^2}.
\end{align}

The quantum complexity is given by the distance of these fluctuations with respect to the ones that are characteristic of $n$-qubit 
Haar-random pure states. As a matter of fact, the Haar-$n$ curve provides a lower limit for universal gate sets \cite{major3}, 
being a reference for identifying quantum complexity unreachable by means of classical computations in the large $n$ limit. 
This criterion not only allows to single out the complexity associated to universal and non-universal classes of random quantum 
circuits, but also of some non-universal but not classically efficiently simulatable ones. Very interesting 
applications in reservoir quantum computing have recently been reported \cite{qrc1,qrc2}. 

\subsection{\label{avr_entanglement}Average Entanglement}

To quantify the entanglement generated by random quantum circuits, we consider the Meyer-Wallach multipartite entanglement measure \cite{meyer2002global}. This quantity was first defined in terms of the wedge product and later Brennen obtained a decomposition in terms of the linear entropy entanglement quantifier, $S_L(\sigma_{A})=1-\Tr(\sigma_A^2)$ \cite{meyer_wallach_brennen, linear_entropy}, being $\sigma_{AB}$ a bipartite system with subsystems $A$, $B$, and $\sigma_A=\Tr_B(\sigma_{AB})$ the reduced state of system $A$. For a pure state of $n$ qubits, $\ket{\psi}$, it is calculated as
\begin{equation}
\label{eq:meyer-wallach}
    Q(\ket{\psi})=\frac{2}{n}\sum_{k=1}^nS_L(\rho_k)=2\left[1-\frac{1}{n}\sum_{k=1}^n\Tr(\rho_k^2)\right],
\end{equation}
being $\rho_k=\Tr_{\hat{k}}(\ketbra{\psi}{\psi})$ the reduced state obtained by tracing out all the qubits, but the $k-$th. This entanglement quantifier is then based on the mean value of the linear entropy considering every possible bipartition one qubit-rest of the system. This quantifier will be maximum when the linear entropy of every bipartition is maximum, i.e., equal to 1. In Eq. \eqref{eq:meyer-wallach}, this is the same as every reduced state being maximally mixed. There are generalizations of this measure for systems with different bipartition sizes (e.g., two qubits and the rest of the system) and different subsystems dimensions \cite{Scott_entanglement_measure, Rigolin_entanglement_measure1, Rigolin_entanglement_measure2} that we are not going to consider in this work.

The circuits will generate very different states depending on the parameter vector given as input. For example, a circuit where all the angles in the parameter vector are $0$ and the input state is $\ket{0}^{\otimes n}$, $n$ number of qubits, will not generate entanglement if only $RX$, $RY$ and $RZ$ parameterized local operations and CNOT gates are performed. This way, to verify the properties of the entanglement generation of the circuit, we perform an average over an ensemble of parameter vectors as
\begin{equation}
    \langle Q \rangle_{\Theta} = \frac{1}{\mathcal{N}}\sum_{i=1}^{\mathcal{N}}Q\left(\ket{\psi(\boldsymbol{\theta}_i)}\right),
\end{equation}
being $\mathcal{N}=10^4$ the sample size and $\boldsymbol{\theta}_i$ different parameter vectors. Each parameter in the parameter vector is sampled according to the uniform distribution between $0$ and $2\pi$. We have also computed the standard deviation, using a very similar procedure where the average and the squared values average were calculated to determine the values. 

The mean value and the standard deviation of this entanglement measure in the case of the Circular Unitary Ensemble (CUE), matching the mean for a uniform distribution of unitaries in the unitary space, was calculated in Refs. \cite{Scott_entanglement_measure, scott_baker_map} and reads, for an $n$-qubits Hilbert space,
\begin{widetext}
\begin{eqnarray}\label{CUEmean}
    \langle Q \rangle_{CUE} &=& \frac{2^n-2}{2^n+1}, \nonumber \\
    \sigma_{CUE}(Q_1) &=& \sqrt{\frac{6(2^n-4)}{(2^n+3)(2^n+2)(2^n+1)n} + \frac{18\cdot2^n}{(2^n+3)(2^n+2)(2^n+1)^2}}.
\end{eqnarray}
\end{widetext}

This relation will be essential to understand the convergence of the entanglement measure as we increase the circuits' number of gates. This convergence is characteristic of random quantum circuits and of circuits that are generating unitary designs of order $2$ \cite{Viola_parameters_of_pseudorandomcircuits, scott_baker_map, Scott_entanglement_measure, correr2024characterizing, express_entang_capab}. This way, it works as a necessary condition of convergence to a $2$-design or to characterize closeness to the generation of uniformly distributed random states. 

\section{\label{sec:results}Results}

\begin{figure}[htb]
\centering
\subfloat[4 qubits]{\includegraphics[width = \columnwidth,keepaspectratio]{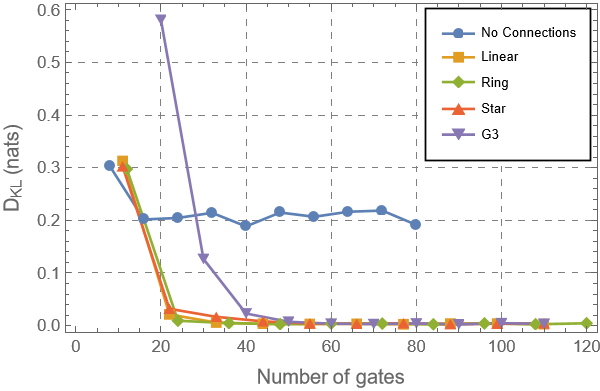}}

\subfloat[8 qubits]{\includegraphics[width = \columnwidth,keepaspectratio]{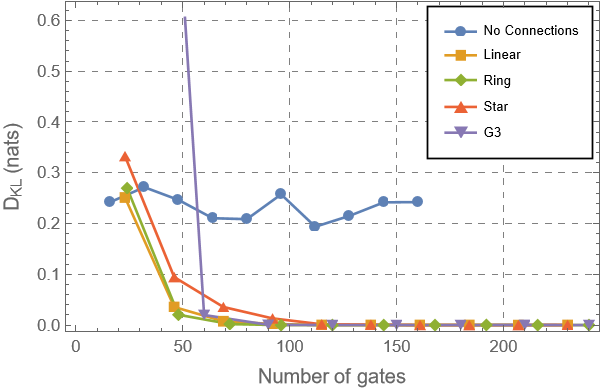}} 

\caption{Expressibility for the different circuits considered where the dimensions are $n=4, 8$ qubits, as a function of the number of gates in the circuit.}
\label{fig:expressibility}
\end{figure}

We have computed the expressibility, fluctuations of the Lorenz curves and the mean entanglement of all circuit classes for different number of 
layers/gates applied. In Fig.~\ref{fig:expressibility} we can see that the expressibility decays very fast to its asymptotic value near zero for all circuits with the exception of the no connections topology for the parameterized classes. This is reasonable since in this case there is no possibility of uniformly distributing the states generated given that non-connected qubits can only provide a strong restriction over all the possibilities of the corresponding state space. But the most interesting feature is that the rest of the parameterized circuits converge faster than the G3 ones to the uniform distribution. The limit is reached at approximately half the number of 
applied gates for the former in the 4 qubit case and at one fourth in the 8 qubit case. In this 8 qubit scenario the star topology performs worse than the G3. It also seems that the behavior of the G3 circuits gets better with a growing qubit number. 

\begin{figure*}[htb]
\centering
\subfloat[4 qubits, 4 layers]{\includegraphics[width = 8cm,keepaspectratio]{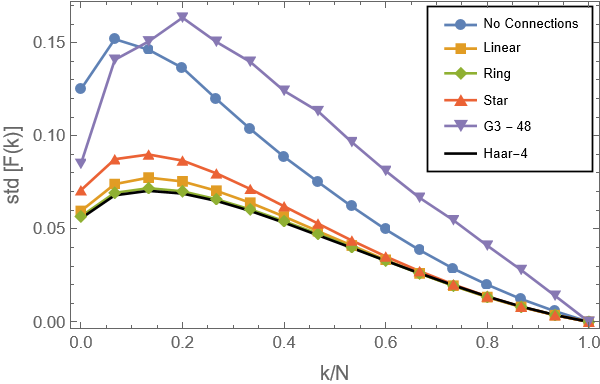}}
\subfloat[8 qubits, 4 layers]{\includegraphics[width = 8cm,keepaspectratio]{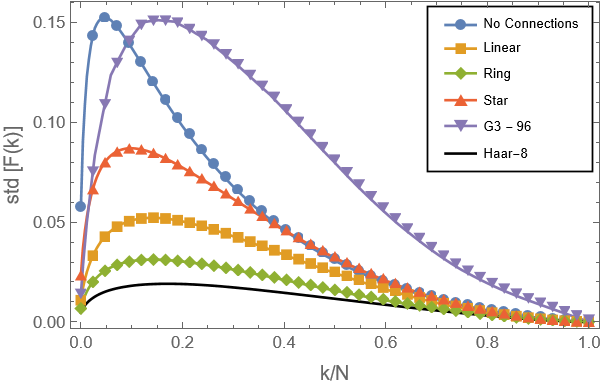}}

\subfloat[4 qubits, 6 layers]{\includegraphics[width = 8cm,keepaspectratio]{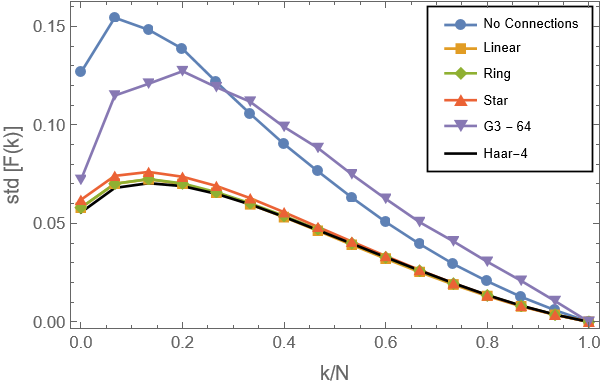}}
\subfloat[8 qubits, 6 layers]{\includegraphics[width = 8cm,keepaspectratio]{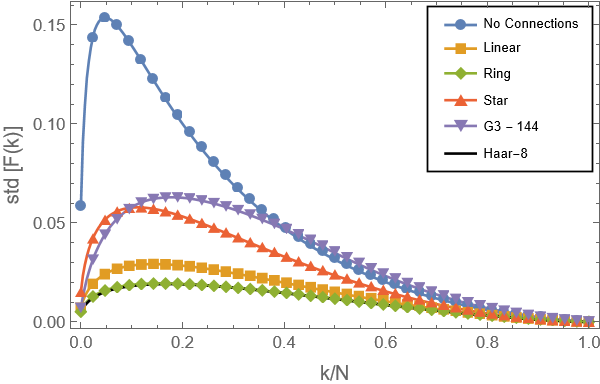}}

\subfloat[4 qubits, 8 layers]{\includegraphics[width = 8cm,keepaspectratio]{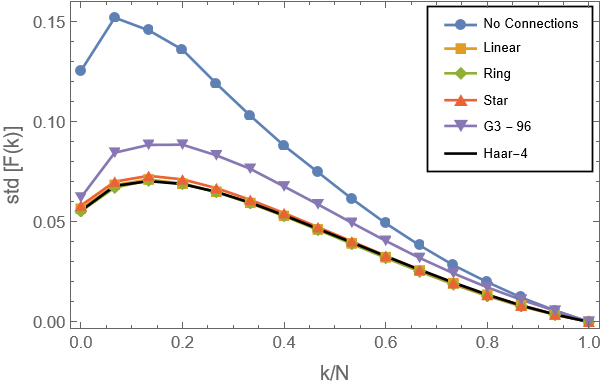}} 
\subfloat[8 qubits, 8 layers]{\includegraphics[width = 8cm,keepaspectratio]{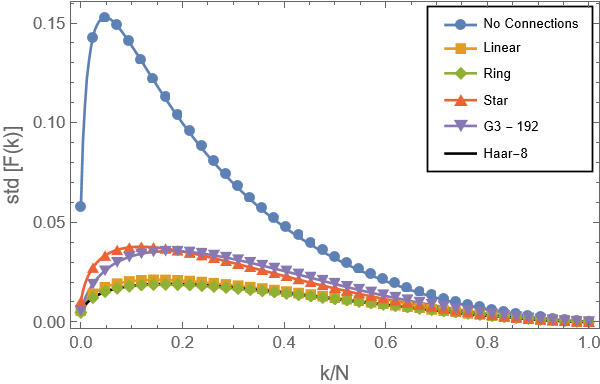}} 

\caption{Fluctuations of the Lorenz curves for the different circuits considered and for Haar sampled unitaries of $n=4, 8$ qubits. The number next to G3 indicates the number of gates applied in the random circuit, which is of the same order of the PQCs for comparison.}
\label{fig:lorenz_curves}
\end{figure*}

But, how can we go deeper into the details of the rate at which 
the states generated reach the optimal complexity? The majorization criterion, based on the fluctuations of the Lorenz curves provides a complementary point of view since it is related to a more general concept than entropy (this latter being at the foundations of the expressibility measure). In Fig. ~\ref{fig:lorenz_curves} we can see that for the 4 qubit case the parameterized circuits with the exception of the no connections scenario, are all near the Haar-4 fluctuations (i.e. the fluctuations corresponding to a uniform sampling over 4 qubits) at 4 layers/48 gates. In fact, the linear and ring topologies have almost converged to this result, while the star shaped circuits have not. The G3 behave almost like the non-connected circuits, and this is so until the bottom panel of the first column corresponding to 8 layers/96 gates where G3 is nearer the Haar-4 results but it has still not converged. This is a remarkable result, but what happens for a growing number of qubits (the behavior of the expressibility of G3 is better)? The case for 8 qubits is shown in the right column of the same figure, where it can be clearly seen that the G3 does not show signs of improvement. It is worth noticing that the parameterized circuits also need more layers/gates to reach the Haar-8 behavior, but eventually they do it at around 6 and 8 layers for the ring and linear topologies, respectively.

\begin{figure*}[htb]
\centering

\subfloat[4 qubits]{\includegraphics[width = 8cm,keepaspectratio]{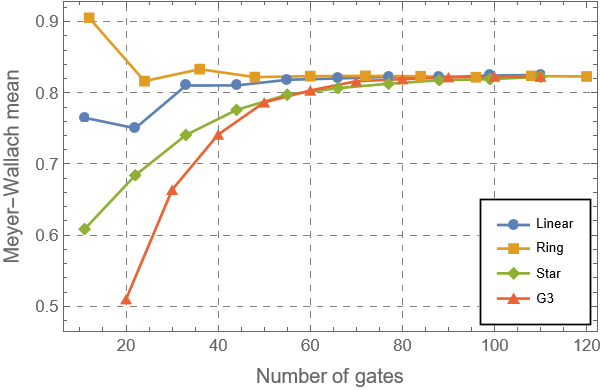}}
\subfloat[8 qubits]{\includegraphics[width = 8cm,keepaspectratio]{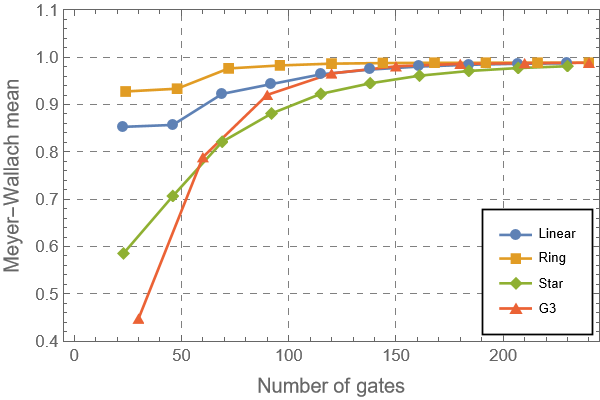}} 

\subfloat[4 qubits]{\includegraphics[width = 8cm,keepaspectratio]{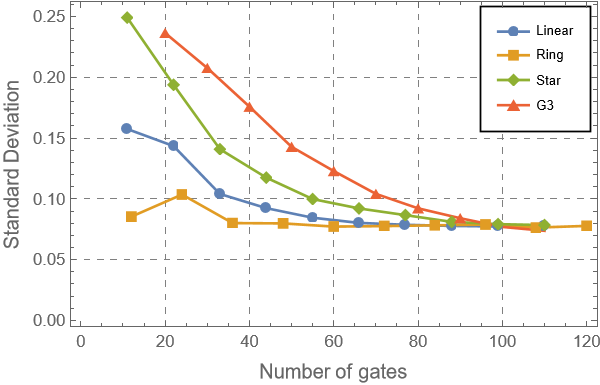}}
\subfloat[8 qubits]{\includegraphics[width = 8cm,keepaspectratio]{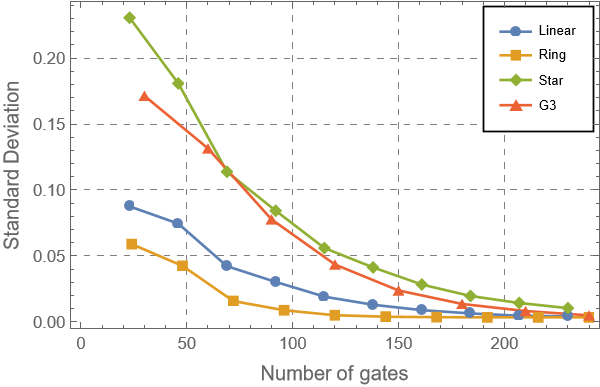}} 
    
\caption{Entanglement and its standard deviation for the different circuits considered for the dimensions of $n=4, 8$ qubits, as a function of the number of gates in the circuit. The only exception not shown is the No Connections circuit, as it does not generate entanglement.}
\label{fig:entanglement}
\end{figure*}

Finally, it is interesting to compare these results with the rate at 
which the mean entanglement is generated. The behavior of this measure 
is qualitatively similar to the expressibility. As a matter of fact, all parameterized circuits outpace G3 ones for the 4 quits case. The CUE limit is reached at around 50 gates for the ring and linear topologies while the star and G3 do it at around 80 gates, almost the double. Looking at the standard deviation, Fig. \ref{fig:entanglement} (c), we can see that the PQCs will always present smaller standard deviations closer to the CUE case than the G3 circuit. In fact, the PQCs present both entanglement values and standard deviations that are closer to the CUE values when compared to G3, from the small number of gates regime to the convergence. For the 8 qubits scenario the situation changes in the same way it changed for the expressibility but slightly better for the G3 circuits, since they reach the CUE limit at approximately 150 gates together with the linear and ring parameterized circuits, while the star topology performs worse, needing about 200 gates for the same result. The standard deviation follows a similar pattern, with the ring and linear circuits presenting values closer to the CUE limit from the beginning, and the star/G3 circuits generating higher values of standard deviations, evolving closer to each other. 

These results are related to similar conclusions made by the authors in a previous work, Ref. \cite{correr2024characterizing}: Random circuits generating entanglement standard deviations closer to the CUE values present a faster increase of the complexity as a function of the number of gates. This statement was made regarding random circuits consisting of PQCs, however the results here indicate that the same can be observed for random circuits consisting of a few generators that are stochastically sampled and applied to the initial state. However, it must be mentioned that this is not a sufficient condition for faster convergence, as there should be enough variability in the circuit structure to achieve a wider range of states.

\section{\label{sec:conclusion}Conclusions}

We have found that parameterized quantum circuits, which have a central role for the development of quantum machine learning among other areas, reach maximum complexity with a fewer number of layers/gates than a paradigmatic class of random quantum circuits generated with H, T and CNOT gates, the G3. The topology of the former circuits is crucial since the less connected have a worse or similar performance than the G3 ones. In fact a linear and specially a ring shape give the most efficient behavior. 

We have used the expressibility and the mean entanglement, which are based on the entropy concept, as measures to characterize such complexity. We have also considered the fluctuations of the Lorenz curves, criterion based on majorization, a more basic concept than entropy (related to a stronger version of the second law of thermodynamics, for example \cite{mixing}). The first two led to similar results where the G3 circuits approached the asymptotic values of complexity at a slower pace than the parameterized ones with the exception of the less connected topologies (no-connections and star). But the last measure not only agreed with the other two, it has shown that the parameterized circuits, in particular the ring topology has an excellent performance, compared to G3.

The advantage of random circuits based on PQCs when compared to random circuits consisting of stochastically applied generators is interesting in the near term quantum computing context. This is owed to the fact that random PQCs have a fixed structure and even with connectivities that are trivial in quantum computers, e.g. the linear topology, they can present a fast increase of complexity by classically sampling parameters of quantum gates.

These results pave the way to try and implement this sort of circuits 
not only in quantum machine learning architectures, but also in the more specific case of quantum reservoir computing and in order to prove quantum supremacy with less resources. For the future we envisage a more realistic evaluation by considering noise, case which has recently led to a very interesting result in terms of its characterization by means of the spectral properties \cite{future}.

\begin{acknowledgments}

This study was financed in part by the Coordenação de Aperfeiçoamento de Pessoal de Nível Superior – Brazil (CAPES) – Finance Code 001, by the Conselho Nacional de Desenvolvimento Científico e Tecnológico - Brazil (CNPq - Grant No. 160851/2021-1 and Grant No. 304891/2022-3), Brazilian National Institute of Science and Technology of Quantum Information (INCT/IQ), and São Paulo Research Foundation - FAPESP (Grant No. 2017/03727-0). Support from CONICET is gratefully acknowledged.

\end{acknowledgments}


%

\end{document}